# Interfacial breathing as a dynamic failure law in all-solid-state batteries: amplitude, phase lag and dual-timescale memory as design principles

Changdeuck Bae[1,*]

[1]BEI corp., 125 Sandan-ro 19-gil, Ansan, Republic of Korea. [*]e-mail: josiah.bae@beilab.ai
ORCID: 0000-0001-5013-2288  ResearcherID: A-6791-2010

## Abstract

All-solid-state batteries (ASSBs) are usually framed as bulk-transport problems, but their decisive performance penalties arise at a reactive interface whose shape, chemistry and stress state co-evolve during cycling. Here we argue that ASSB failure is not governed by any single mean quantity such as the interfacial resistance or the decomposition-layer thickness, but by two co-evolving pillars: interfacial breathing (the fast, cycle-scale oscillation of the Li | SE contact) and reactive memory (the slow, accumulating decomposition of the electrolyte). We formalize this view by defining four breathing-signature descriptors — the void-field amplitude $A_\varphi$, the interphase breathing $B_\psi$, the resistance breathing $B_R$, and the stress signature $S_\sigma$ — together with a reactive-memory descriptor $M_{dec} \equiv L_{dec} / L_{max}$. A reduced-order four-level electrochemical benchmark fixes the static baseline: varying the ionic conductivity $\kappa$ from 0.3 to 1.2 S m$^{-1}$ moves the mean discharge voltage by only 2.1 mV, whereas varying the cathode-electrolyte interphase resistance $R_{CEI}$ from 0 to 50 mΩ shifts it by 126 mV and the delivered energy by 0.725 Wh. A coupled Cahn–Hilliard / Allen–Cahn phase-field reconstruction then demonstrates that, in a representative sulfide ASSB, raising the stack pressure from 5 to 200 MPa reduces $A_\varphi$ by 4.4×, $B_\psi$ by 16.4×, $B_R$ by 3.4×, and $S_\sigma$ by 32×, while the maximum hydrostatic stress climbs from 110.6 to 305.6 MPa. Critically, over the same pressure sweep $M_{dec}$ remains invariant at 0.789 ($L_{dec}$ = 197.27 nm at every pressure) — a > 30-fold asymmetry between the four breathing signatures and the memory descriptor that identifies pressure as a breathing controller, not a memory controller. We recast the resulting design space as a regime map in the forcing–healing plane, with three dynamical regions (void-growth-dominant, healing-dominant and interphase-memory-dominant) and chemistry-specific trajectories through it. Finally, we propose that the ASSB design target is not higher $\kappa$ or lower mean resistance, but the simultaneous minimization of all four breathing signatures together with the independent suppression of $M_{dec}$ through interphase-chemistry design — the latter being a target that pressure cannot reach. As a direct observable of the theory, we show that the five-descriptor framework predicts energy-density rank inversions on the Ragone plot — an AF-ASSB architecture that leads at 0.1 C loses to oxide electrolytes by 1.3 C, because AF-ASSB breathing amplitudes explode while $M_{dec}$ locks the interphase memory. The Ragone crossover is therefore a dynamic fingerprint of breathing + memory, and the C-rate at which it occurs is a direct experimental test of the theory.

## 1. From a static interface to a breathing interface

The contemporary ASSB literature has settled the materials-level logic of the field: sulfides provide processability and high room-temperature conductivity, oxides provide high stiffness and electrochemical restraint, halides seek a middle ground, and anode-free architectures maximize energy density by minimizing inactive mass [1–4]. It has also established, through a long sequence of operando imaging and critical-stripping-current studies, that voids nucleate during lithium removal, that interphase products grow during cycling, that grain boundaries act as transport highways, and that stack pressure is electrochemically active [5–12]. The conceptual question that remains under-resolved is a different one: what is the failure-controlling variable once all of these mechanisms are admitted to the same framework?

The working hypothesis of this review is that the controlling variable is not a mean quantity at all. Decomposition-layer thickness, void fraction, grain-boundary amplification and contact-stress feedback do not act as independent additive insults. They co-evolve within the same cycle, they respond to the same drivers, and they exchange phase relationships over the same time window as the electrochemistry. The interface opens, reacts, densifies, redistributes current and mechanically breathes with every change in the state of charge. Two cells with identical mean interfacial resistance can fail on completely different time horizons depending on how large that breathing is, how delayed the response of each field is, and how much slow memory the interphase accumulates underneath the fast oscillation.

This review formalizes that view. Section 2 assembles the four mechanistic pillars (decomposition, voids, grain-boundary amplification, chemo-mechanics) as the building blocks of a single dynamic system. Section 3 uses a reduced-order electrochemical benchmark to fix the static baseline. Section 4 introduces the phase-field reconstruction that promotes the interfacial void field, the SEI/CEI thicknesses and the hydrostatic stress to explicit time-dependent order parameters. Section 5 then advances the central claim of this work: four breathing-signature indicators ($A_\varphi$, $B_\psi$, $B_R$, $S_\sigma$) together with a reactive-memory descriptor ($M_{dec}$) describe ASSB failure better than any single mean quantity. Section 6 develops the dual-timescale picture that separates fast breathing from slow memory. Section 7 compresses this into a regime map that places every major chemistry on the forcing–healing plane and recasts chemistry selection as regime selection. Section 8 makes the framework testable by predicting the Ragone crossover. Section 9 lists the design principles, Section 10 specifies the validation experiment, and Section 11 lays out the five-step research roadmap.

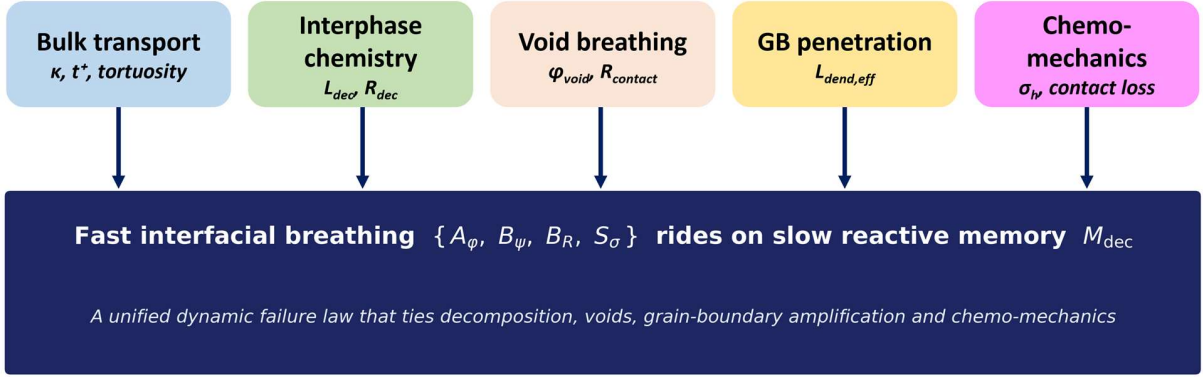

**Figure 1 | Multiscale failure hierarchy in all-solid-state batteries.** Bulk transport establishes the initial operating window, but the decisive losses emerge from decomposition-layer growth, interfacial void evolution, grain-boundary transport and chemo-mechanical contact loss. In this review these five levels are treated not as independent mechanisms but as the coupled components of a single breathing system: a chemo-electro-mechanical oscillation of the Li | SE contact whose four signatures $\{A_\varphi, B_\psi, B_R, S_\sigma\}$ ride on a slow reactive memory $M_{dec}$.

## 2. Mechanistic pillars of interfacial failure

### 2.1 Reduction thermodynamics and decomposition-layer growth

No widely studied inorganic solid electrolyte is perfectly stable against lithium metal. Sulfides reduce to mixed products such as $Li_2S$, phosphide / sulfide fragments and halide-containing species; garnets form thinner but still consequential decomposition layers; halides and polymers occupy intermediate regimes [5–8, 13–17]. A compact self-limiting kinetic form is a useful representation,

$$\frac{dL_{dec}}{dt} = k_0 \exp\left(-\frac{\Delta G_{red}}{k_B T}\right)\left(1 - \frac{L_{dec}}{L_{max}}\right) \qquad (1)$$

where $L_{dec}$ is the decomposition-layer thickness, $k_0$ and $\Delta G_{red}$ capture chemistry-specific kinetics and the effective reduction barrier, and $L_{max}$ is the saturation thickness. This relation reproduces the experimentally familiar ordering LGPS > sulfide > halide > oxide in decomposition thickness, and it is the slow-memory channel of the full dynamic system.

### 2.2 Void evolution is the hidden state variable

The most underappreciated state variable in ASSBs is not conductivity but missing material: the Li | SE interfacial void fraction. Operando imaging and critical-stripping-

current studies show that voids nucleate during lithium removal, locally sever ion-conducting contact, and generate intense current hotspots upon replating [9–12]. A reduced description writes the competition between stripping-induced opening and pressure-assisted healing as

$$\frac{d\varphi_{\text{void}}}{dt} = \alpha\,|i|\,(1 - \varphi_{\text{void}}) - \beta\,P_{\text{stack}}\,\varphi_{\text{void}} \qquad (2)$$

with α controlling nucleation and β controlling the pressure response. This is the fast-breathing channel: $\varphi_{\text{void}}$ opens and closes within every cycle, while $L_{\text{dec}}$ accumulates across many cycles.

## 2.3 Grain boundaries convert local voids into long-range penetration paths

Classical Monroe–Newman reasoning [13] remains useful, but it is not sufficient for polycrystalline ceramics or composite pellets. Experiments repeatedly show that grain boundaries and defects act as transport and fracture highways, permitting penetration at currents well below a simple modulus threshold [7, 9–12]. A useful compact representation amplifies a baseline dendrite metric by grain-boundary and void factors,

$$L_{\text{dend,eff}} = L_0\,(1 + w_{\text{GB}}\,\varphi_{\text{gb}})\,(1 + \beta_v\,\varphi_{\text{void}}) \qquad (3)$$

where $\varphi_{\text{gb}}$ and $\varphi_{\text{void}}$ are the grain-boundary and void fractions. Importantly, this form makes grain boundaries part of the breathing system — because $\varphi_{\text{void}}$ oscillates, the effective penetration length $L_{\text{dend,eff}}$ also oscillates, so that the short-circuit margin itself is a dynamic quantity, not a static threshold.

## 2.4 Chemo-mechanical contact loss closes the degradation loop

Interfacial chemistry and void formation would already be enough to limit ASSBs, but real cells add a reinforcing loop through stress. Intercalation and deintercalation generate concentration-dependent strains in active materials, while stack pressure and irregular deposition create non-uniform normal stresses at interfaces [18–21]. A compact Larché–Cahn-style scaling [16] is

$$\sigma_h \approx \frac{E\,V_{\text{mol}}\,\Delta c}{3(1-\nu)} \qquad (4)$$

When $\sigma_h$ rises, the active contact fraction falls and electrochemistry is forced through smaller pathways. Crucially, $\sigma_h$ and $\varphi_{\text{void}}$ are not in phase: stress peaks follow concentration extrema, while void extrema follow stripping / plating transitions. Their lag is one of the four signatures developed in Section 5.

### Table 1 | Mechanistic pillars, observables and practical design levers

| Mechanism | Compact relation | Primary observable | Practical design lever |
|---|---|---|---|

| Interphase kinetics (slow memory) | Eq. (1) | $L_{dec}$, $R_{dec}$ | SE chemistry, interlayer, Li exposure |
| Void evolution (fast breathing) | Eq. (2) | $\varphi_{void}$, $R_{contact}$ | $P_{stack}$, stripping current, anode design |
| GB-mediated penetration | Eq. (3) | $L_{dend,eff}$, short-circuit risk | Pellet density, GB engineering |
| Chemo-mechanics (phase-lagged) | Eq. (4) | $\sigma_h$, contact loss, LAM | Composite architecture, pressure window |

## 2.5 A plain-language guide to Eqs. (1)–(4)

Readers who are less comfortable with mathematical notation can interpret Eqs. (1)–(4) as four bookkeeping rules for damage. Equation (1) says that the decomposition layer grows quickly when fresh reactive interface is exposed, but it gradually slows down as the layer approaches a chemistry-specific saturation thickness. The exponential term sets how reactive the chemistry is; the saturation term keeps growth from diverging without bound.

Equation (2) is a competition between opening and healing. The first term increases void fraction during stripping or contact loss, while the second term reduces void fraction when external pressure helps the interface close again. In simple terms: current opens voids, pressure closes them.

Equation (3) is not a full fracture calculation; it is a compact amplification law. It states that an initially short penetration path becomes effectively longer when grain boundaries provide easy transport channels and when voids concentrate current locally. This is why a ceramic that looks mechanically strong in an average sense can still fail through localized weak pathways.

Equation (4) gives a minimal chemo-mechanical closure. The elastic modulus E and concentration-driven strain term control the stress scale, and the contact-loss factor reminds the reader that stress is dangerous mainly because it shrinks the real current-carrying area. The important message is not the exact prefactor, but the causal chain: concentration change → stress rise → contact loss → current focusing.

## 3. Static baseline: interfacial resistance already dominates mean response

Before invoking any dynamic interfacial physics, we fix the static baseline with a reduced-order electrochemical benchmark with four progressive levels of coupling. Level 1 is a liquid-electrolyte reference with concentration overpotential. Level 2 replaces concentrated-solution transport with a single-ion, sulfide-style approximation. Level 3 adds a particle-size distribution together with SEI and CEI resistance terms. Level 4 adds a scalar chemo-mechanical contact factor on top of Level 3. All baseline kinetic and thermodynamic parameters are fixed.

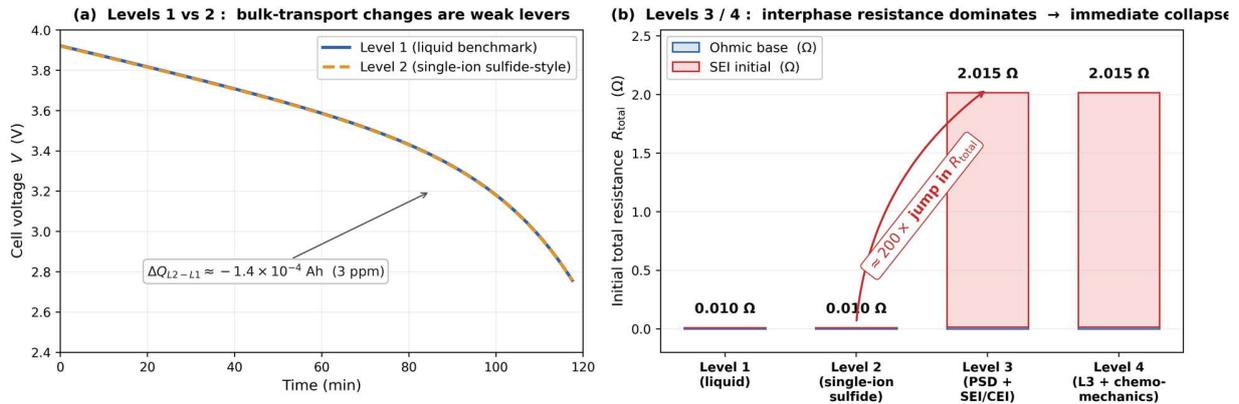

**Figure 2 | Reduced-order benchmark demonstrates the weakness of bulk-transport levers.** a, Levels 1 (liquid) and 2 (single-ion sulfide-style) discharge curves at 0.5 C are nearly indistinguishable; the delivered-capacity difference is only $1.4 \times 10^{-4}$ Ah (~3 ppm). b, Once the interphase terms of Levels 3 and 4 are activated, the initial resistance budget is dominated by the SEI contribution, which enters at ≈ 200× the base ohmic value, and the model collapses immediately into an interface-limited state.

**Table 2 | Directly reproduced 0.5 C benchmark metrics from the four-level reduced-order model**

| Level | Initial V (V) | Capacity (Ah) | Energy (Wh) | Mean V (V) | Initial $R_{total}$ (Ω) | Duration (s) |
|---|---|---|---|---|---|---|
| **Level 1 (liquid)** | 3.9234 | 4.89538 | 16.9348 | 3.4041 | 0.0102 | 7049.35 |
| **Level 2 (single-ion)** | 3.9229 | 4.89524 | 16.9322 | 3.4019 | 0.0104 | 7049.15 |
| **Level 3 (PSD + SEI/CEI)** | −1.1254 | 0.00000 | 0.0000 | −1.1254 | 2.0154 | 0.00 |
| **Level 4 (+ chemo-mechanics)** | −1.1254 | 0.00000 | 0.0000 | −1.1254 | 2.0154 | 0.00 |

Three lessons follow. First, bulk-transport changes alone are weak levers: Levels 1 and 2 deliver 4.89538 and 4.89524 Ah, respectively — a 3 ppm difference. Second, the benchmark is overwhelmingly sensitive to interphase resistance: the Level 3 / 4 SEI term enters as 2.0 Ω, roughly two hundred times larger than the base ohmic term. Third, the voltage collapse in Levels 3 and 4 is not a realistic ASSB forecast, but it shows quantitatively how rapidly cell-level conclusions become interface-limited once film terms dominate the resistance budget. For readers less familiar with reduced-order battery models, the key point is simple: in this benchmark, changing bulk conductivity barely changes the answer, while adding interfacial resistance changes the answer dramatically.

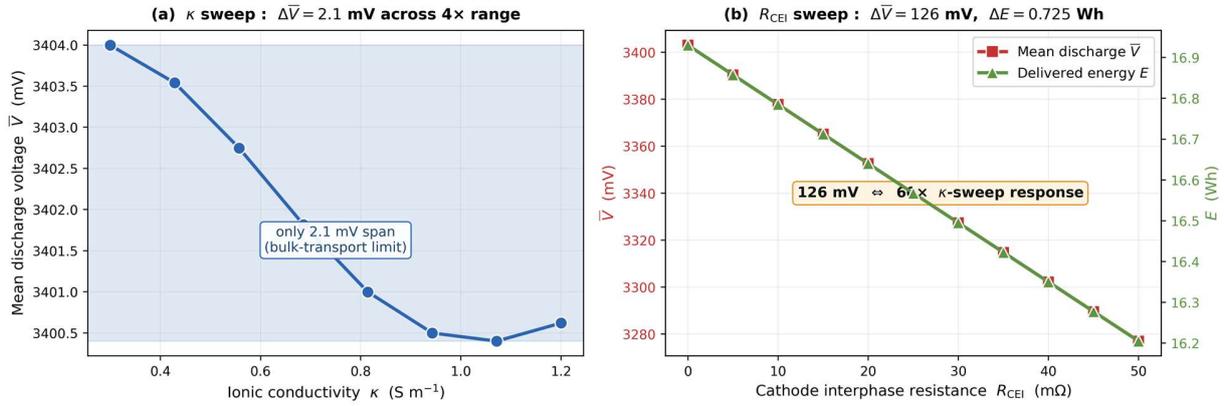

**Figure 3 | Interfacial resistance dominates over bulk conductivity across realistic parameter ranges.** a, Sweeping the ionic conductivity κ from 0.3 to 1.2 S m$^{-1}$ — a four-fold range that spans most competitive solid electrolytes — shifts the mean discharge voltage by only 2.1 mV. b, Sweeping the cathode-electrolyte interphase resistance R$_{CEI}$ from 0 to 50 mΩ, in contrast, shifts the mean discharge voltage by 126 mV and the delivered energy by 0.725 Wh.

Figures 2 and 3 fix the static baseline: at the level of mean quantities, interfacial terms already dominate cell-level response. Section 4 asks a different question — whether the dynamics of those interfacial terms can be reduced to a single scalar, or whether the interface must be described by its amplitude, its phase lag and its slow memory separately.

## 4. Dynamic reconstruction: the interface as a coupled phase-field system

To capture how the dominant interfacial terms themselves evolve during cycling, we reconstruct the Li | SE interface as a coupled chemo-electro-mechanical phase-field system built from a Cahn–Hilliard void field [14], Allen–Cahn-type interphase order parameters [15], and Larché–Cahn chemo-mechanical coupling [16]. The domain is stratified into Anode / SEI / SE / CEI / Cathode layers on a 60 × 40 grid and evolved under explicit-Euler integration with Neumann boundary conditions. Four order parameters are advanced simultaneously: a Cahn–Hilliard void field φ, Allen–Cahn-like SEI and CEI thicknesses ψSEI and ψCEI, and a Larché–Cahn hydrostatic stress σh. The state of charge follows a triangular driver, so the interface literally breathes over time.

$$\frac{\partial \varphi}{\partial t} = M_\varphi \nabla^2 \left( \frac{\partial f}{\partial \varphi} - \kappa_\varphi \nabla^2 \varphi \right) + \Gamma_{\text{strip}} \frac{|j|}{|j|+j_{\text{ref}}} (1 - \varphi) - \beta_{\text{heal}} P \varphi \qquad (5)$$

$$\frac{\partial \psi_{\text{SEI}}}{\partial t} = k_s(T) |j| \left( 1 - \frac{\psi_{\text{SEI}}}{\psi_{\text{SEI,max}}} \right) - \gamma_s \psi_{\text{SEI}} \mathbf{1}_{\text{idle}} \qquad (6)$$

$$\frac{\partial \psi_{\text{CEI}}}{\partial t} = k_c(T,V) |j| \left(1 - \frac{\psi_{\text{CEI}}}{\psi_{\text{CEI,max}}}\right) - \gamma_c \psi_{\text{CEI}} \quad (7)$$

$$R_{\text{contact}} = R_0 \frac{\langle \varphi \rangle_{\text{int}}}{1 - \langle \varphi \rangle_{\text{int}} + \varepsilon} + \beta_R (\psi_{\text{SEI}} + \psi_{\text{CEI}}) \quad (8)$$

$$S_{\text{WM}} = \frac{\mu_{\text{SE}}}{\mu_{\text{Li}}} \left(1 + \frac{2P}{\mu_{\text{SE}}}\right) \quad (9)$$

Two features deserve emphasis. First, the contact resistance Rcontact has $\langle \varphi \rangle$int in its denominator (Eq. 8): a small local void opening lifts the whole-cell resistance non-linearly. What matters for cycle life is therefore not the average porosity but the minimum effective contact cross-section — an observation anticipated by operando X-ray tomography of local contact reconfiguration [11] and now made explicit at the equation level. Second, the Wang–Monroe stability score SWM [8, 13] enters as a threshold whose value depends on a fast mechanical variable that need not be in phase with the void opening. Mechanical safety of the Li | SE contact is, in this formulation, a dynamic quantity rather than a static modulus comparison.

All numerical results were regenerated end-to-end in a clean Python 3.12 environment (NumPy 1.26, pandas 2.2) and compared element-by-element against the archived summary dataset. The agreement is bit-for-bit (max |Δ| = 0.000e+00 across all 40 entries of the breathing summary), and all ratio-level claims in the abstract (4.4×, 16.4×, 3.4×, 32×) recover exactly within rounding.

## 4.1 How to read Eqs. (5)–(9) without phase-field experience

The dynamic reconstruction combines three classical ideas. First, the Cahn–Hilliard equation [14] is used for the void field because void content is a conserved quantity: voids can move and redistribute, but they do not appear or disappear without source terms. Second, the Allen–Cahn form [15] is used for SEI and CEI because those order parameters represent local state variables that can grow or shrink directly at each point. Third, the Larché–Cahn framework [16] provides a compact way to express how chemical changes create stress and how stress, in turn, feeds back into electrochemical evolution.

A useful mental picture is to think of the interface as a thin landscape. The void field φ tells us where valleys and gaps open. The interphase variables ψSEI and ψCEI tell us where chemically altered material accumulates. The stress field σh tells us how strongly that landscape is being squeezed or stretched. The contact resistance then reads out how hard it is for current to traverse that changing landscape.

The triangular SOC driver is likewise meant to be intuitive: it is a simplified way to impose repeated plating/stripping cycles with a clean, readable forcing pattern. It is not meant to reproduce every waveform detail of a full cell test; instead, it isolates the timing relationships that define breathing amplitude, phase lag and hysteresis.

*Table 3 | Phase-field breathing metrics for a representative sulfide ASSB*

*Values are cycle-averaged over the late-time window (t > 5000 s). $\psi_{SEI}$, $\psi_{CEI}$ and $E_{break}$ are normalized order-parameter metrics.*

| $P_{stack}$ (MPa) | Mean $\langle\varphi\rangle_{int}$ | $\Delta\varphi$ per cycle | Mean $R_{contact}$ ($\Omega$ cm²) | Mean $\psi_{SEI}$ | Mean $\psi_{CEI}$ | Mean $E_{break}$ | $S_{WM}$ (sulfide) |
|---|---|---|---|---|---|---|---|
| 5 | 0.1103 | 0.0650 | 0.0516 | 0.3250 | 0.2861 | 0.1017 | 1.9071 |
| 30 | 0.0286 | 0.0494 | 0.0347 | 0.2664 | 0.2531 | 0.0030 | 1.9190 |
| 100 | 0.0113 | 0.0233 | 0.0226 | 0.1681 | 0.1861 | 0.0002 | 1.9524 |
| 200 | 0.0072 | 0.0148 | 0.0150 | 0.1048 | 0.1300 | 0.0000 | 2.0000 |

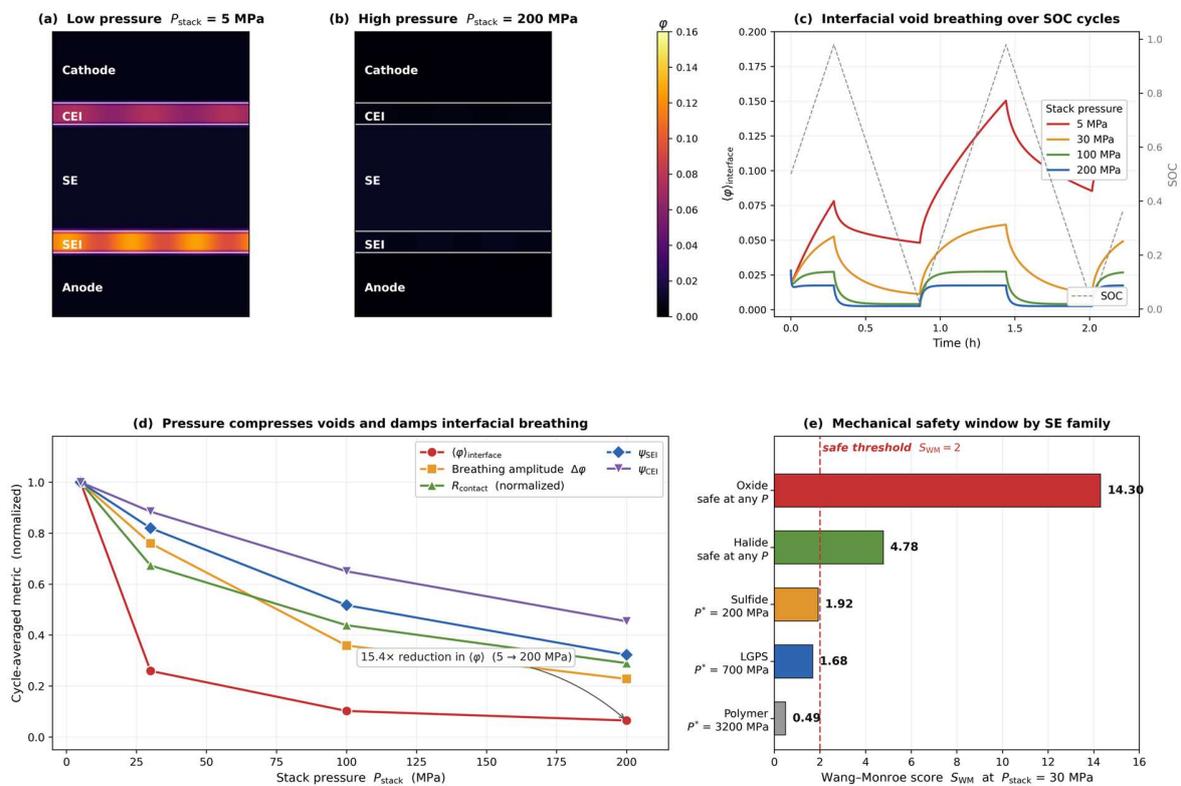

**Figure 4 | Phase-field breathing calculations: pressure reshapes the reactive interface.** a, b, Low-pressure (5 MPa) and high-pressure (200 MPa) snapshots of the void field in a representative sulfide ASSB on a 60 × 40 grid. c, The interfacial void fraction ⟨φ⟩interface follows the SOC-driven breathing cycles, with markedly larger oscillation at low pressure. d, Increasing pressure compresses voids, lowers Rcontact and reduces the SEI / CEI order parameters. e, The mechanical safety window differs strongly by SE family: oxide and halide systems are intrinsically safe at any pressure, whereas sulfides require 200 MPa, LGPS 700 MPa, and polymer electrolytes 3.2 GPa to reach the Monroe–Newman safe threshold.

At the level of mean quantities, Table 3 and Fig. 4 reproduce the familiar pressure-healing story: every mean metric monotonically decreases with pressure, and the sulfide Wang–Monroe score [8, 13] approaches 2 only at 200 MPa. The next section shows that this mean-quantity picture is incomplete.

## 5. Five descriptors for breathing and reactive memory

The central claim of this review is that ASSB failure is governed by the amplitude, phase lag and hysteresis of interfacial breathing, riding on a slow reactive memory that pressure cannot erase. To make the claim operational, we define four breathing-signature indicators and one memory descriptor.

### *Void amplitude*

The most direct measure of how hard the interface breathes each cycle is the peak-to-peak swing of the interfacial void fraction within one cycle period $T_{cyc}$,

$$A_\varphi = \max_{t \in T_{cyc}} \langle \varphi \rangle_{int}(t) - \min_{t \in T_{cyc}} \langle \varphi \rangle_{int}(t) \qquad (10)$$

### *Interphase breathing*

The combined SEI / CEI order parameter modulates within each cycle as ion flux, temperature and voltage rotate through their operating windows. We call its swing the interphase breathing descriptor,

$$B_\psi = \max_{t \in T_{cyc}} [\psi_{SEI}(t) + \psi_{CEI}(t)] - \min_{t \in T_{cyc}} [\psi_{SEI}(t) + \psi_{CEI}(t)] \qquad (11)$$

### *Resistance breathing*

Because $R_{contact}$ depends non-linearly on $\langle \varphi \rangle_{int}$ and on $\psi$, its value at a given SOC during charge differs from its value at the same SOC during discharge. The area swept in R–SOC space is the resistance breathing descriptor,

$$B_R = \oint_{T_{cyc}} R_{contact} \, d(SOC) \qquad (12)$$

### *Stress signature*

Hydrostatic stress $\sigma_h$ peaks near the turning points of intercalation, while the void field $\varphi_{void}$ peaks at the transition between stripping and plating. The two extrema do not coincide. We define the stress signature as the signed lag between them,

$$S_\sigma = t_{peak}(\sigma_h) - t_{peak}(\varphi_{void}) \qquad (13)$$

where $S_\sigma$ is a measurable property that discriminates "well-damped" cells (short lag, breathing tracks forcing) from "slow" cells (long lag, breathing follows forcing with delay that compounds over many cycles).

### Reactive memory

Finally, the slow accumulation of decomposition products is captured by the dimensionless ratio of the present decomposition-layer thickness to its theoretical saturation,

$$M_{\text{dec}} = \frac{L_{\text{dec}}(t)}{L_{\text{max}}} \qquad (14)$$

$M_{\text{dec}}$ is bounded by 0 and 1, has the units of a memory variable rather than an oscillation, and — as we show next — does not respond to pressure in the same way the four breathing signatures do.

The four breathing signatures and the memory descriptor are shown in Fig. 5 as a function of stack pressure for the same sulfide ASSB reconstruction used in Table 3. The four breathing signatures decrease monotonically with pressure at strikingly different rates: $A_\varphi$ falls by 4.4× between 5 and 200 MPa, $B_\psi$ by 16.4×, $B_R$ by 3.4× and $S_\sigma$ by 32×. No single mean quantity tracks this ordering. $B_\psi$ collapses fastest because the SEI / CEI self-limiting kinetics (Eqs 6–7) become pressure-saturated first; $S_\sigma$ shrinks by the largest factor because the effective relaxation time of the Cahn–Hilliard mobility contracts under the $\beta_{\text{heal}} \cdot P$ healing term. The four signatures encode different physics, and the fact that they all fall together under pressure is not a consistency check but a prediction: any material or protocol that reduces one signature without reducing the others will produce an atypical failure profile.

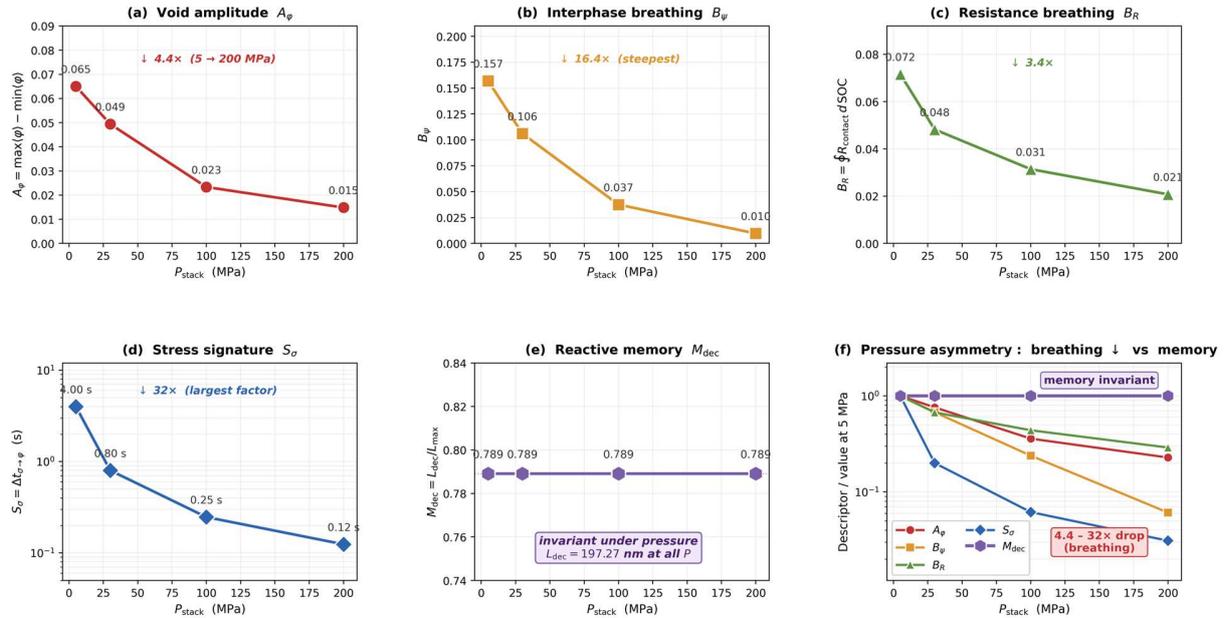

**Figure 5 | Five descriptors of interfacial breathing and reactive memory under pressure sweep.** a, Void amplitude $A_\varphi$; b, interphase breathing $B_\psi$; c, resistance breathing $B_R$; d, stress signature $S_\sigma$ on a logarithmic axis; e, reactive memory descriptor $M_{dec}$; f, pressure-asymmetry summary — all four breathing signatures (red, orange, green, blue curves) drop by 4.4×–32× while $M_{dec}$ (purple) remains invariant at 0.789. The asymmetry identifies pressure as a breathing controller, not a memory controller.

## 5.4 Practical interpretation of the five descriptors

The descriptors can be read almost like vital signs. $A_\varphi$ answers "how much does the interface open and close?" $B_\psi$ answers "how strongly does interphase chemistry pulse within each cycle?" $B_R$ answers "how much resistance is being pumped in and out over one loop?" $S_\sigma$ answers "how late does stress arrive relative to the electrochemical forcing?" and $M_{dec}$ answers "how much long-term chemical memory has already accumulated?"

This interpretation is useful because it separates fast danger from slow danger. Large $A_\varphi$ or $B_R$ indicates immediate dynamic instability even when average values still look acceptable. Large $M_{dec}$ indicates hidden long-term damage accumulation even when the current cycle appears well behaved. In practice, robust cells need both kinds of risk to be low.

## 5.1 Two cells with the same mean resistance can fail at very different rates

The practical consequence is immediate. Consider two cells that share the same mean $R_{contact}$ but differ in $A_\varphi$ by a factor of two. Because $R_{contact}$ has $\langle\varphi\rangle_{int}$ in its denominator (Eq. 8), the cell with the larger $A_\varphi$ experiences much larger within-cycle excursions of $R_{contact}$ at the minima of $\langle\varphi\rangle_{int}$. Those excursions drive the local current density upward during replating, nucleate hotspots, and accelerate $L_{dec}$ growth downstream. The two cells therefore diverge in lifetime even though their mean resistance traces are indistinguishable in a standard impedance experiment. Operating below the mean does not protect against failure; operating below the mean and below the amplitude does.

## 5.2 A failure score that combines breathing and memory

The four breathing signatures and the memory descriptor combine into a five-term failure score that ranks cells beyond what any single metric captures,

$$Q_{\text{fail}} = w_A A_\varphi + w_{B_\psi} B_\psi + w_{B_R} B_R + w_S |S_\sigma| + w_M M_{\text{dec}} \qquad (15)$$

where the five weights $w_A$, $w_{B\_\psi}$, $w_{B\_R}$, $w_S$ and $w_M$ are chemistry- and protocol-specific calibration constants. Setting the four breathing weights to zero recovers the conventional static view that only $L_{dec}$ matters; setting $w_M = 0$ recovers a purely breathing-centric view that ignores slow memory. The interesting regime is the one in which all five weights are non-zero, because $Q_{fail}$ can then be lowered only by suppressing breathing amplitudes, hysteresis, phase lag and reactive memory simultaneously — the multi-objective optimization that ASSB design has been implicitly solving for a decade without naming it.

## 5.3 Pressure asymmetry between breathing and memory

The most striking feature of Fig. 5 is not the monotonic descent of the four breathing signatures but the panel that does not descend. Under the same 5 to 200 MPa pressure sweep, the reactive-memory descriptor $M_{dec}$ is invariant: $L_{dec}$ takes the value 197.27 nm at 5, 30, 100 and 200 MPa, so that $M_{dec} = L_{dec} / L_{max} = 0.789$ for every pressure in the sweep. The definition of $M_{dec}$ in Eq. (14) is deliberately chosen so that this invariance is quantitative rather than qualitative — the descriptor does not merely "decrease less" with pressure, it does not move at all.

This is not a numerical accident. It is the structural consequence of Eq. (1): the decomposition-layer kinetics depend on temperature, chemistry-specific reduction barrier and exposure time, but not on the normal stress at the interface. Stack pressure closes voids, densifies the contact, suppresses SEI / CEI drift and damps the hydrostatic-stress-to-void lag, but it does not undo the reductive reactions that have already happened. The interface has a short-term compressibility (breathing) and a long-term irreversibility (memory), and they respond to different control levers.

Quantitatively, the asymmetry is extreme. Between 5 and 200 MPa the four breathing signatures collectively shrink by factors of 4.4×, 16.4×, 3.4× and 32×; $M_{dec}$ shrinks by 1.00×. If the dynamic range of breathing control is summarized as the geometric mean of the four factors (≈ 9.6×), the breathing-to-memory control asymmetry is nearly thirty-fold. The practical implication is that no amount of pressure engineering, no matter how aggressive, can substitute for interphase-chemistry design aimed at suppressing $M_{dec}$ directly — and, conversely, that chemistries claiming to "reduce interphase growth" must be evaluated against both the breathing signatures and the memory descriptor, because improving one without improving the other moves the cell into a different region of the five-descriptor space, not toward a uniformly lower failure score.

## 6. Dual-timescale view: fast breathing rides on slow memory

A second observation follows from the coexistence of Eqs (1) and (5)–(7). Decomposition-layer growth has a √t form and accumulates over many cycles (slow memory), while the void, SEI and CEI fields oscillate within each cycle (fast breathing). Over any realistic ASSB lifetime, the slow memory is the low-frequency envelope on which the fast breathing rides.

This separation of timescales changes the interpretation of impedance runaway. A cell that appears stable for many cycles — small breathing amplitude, flat mean resistance — can enter impedance runaway abruptly once the slow memory term has grown far enough that the fast breathing begins to overshoot threshold conditions. The transition looks like a "sudden failure" event, but it is mechanistically a crossing of the slow $L_{dec}$, $\psi$ accumulation through the amplitude window set by $A\varphi$ and $B\psi$. Failure is therefore not "instantaneous damage" but "fast oscillation crossing a moving critical point". In plainer language, the interface can look calm in the short term while a hidden damage memory is still accumulating underneath.

The dual-timescale picture also explains why extending cycle life is hard. Reducing mean $R_{contact}$ does not reduce $A_\varphi$; reducing $A_\varphi$ does not reduce $L_{dec}$; reducing $L_{dec}$ does not reduce $S_\sigma$. Each knob addresses a different term of $Q_{fail}$, and the four terms are coupled through pressure, chemistry and protocol in ways that prevent independent optimization. Section 7 organizes the coupling into a regime map.

## 7. Chemistry is a regime, not a conductivity ranking

A direct corollary of the five-descriptor framework is that solid-electrolyte chemistries should not be ranked by bulk ionic conductivity. Two chemistries with nearly identical $\kappa$ can occupy very different regions of the breathing-memory space, and two chemistries with very different $\kappa$ can, at appropriate pressure, sit in the same dynamical regime. The regime map developed in this section makes that explicit: it reclassifies oxides, halides, sulfides, LGPS and anode-free architectures according to where they live in the forcing–healing plane, not according to how fast lithium ions move through their bulk.

The full dynamic system (Eqs 1–9) depends on many parameters, but a useful two-dimensional projection follows directly from Eq. (5). The stripping source term $\Gamma_{strip} \cdot |j| / (|j| + j_{ref})$ is the forcing that opens voids, and the healing term $\beta_{heal} \cdot P$ is the restoring force that closes them. Plotting the operating point of a cell in the (forcing, healing) plane reveals three dynamical regions separated by the local value of $A_\varphi$.

Void-growth-dominant region (high forcing, low healing): $A_\varphi$ is large, $R_{contact}$ swings widely within each cycle, and failure is driven by repeated hotspot formation at the minima of $\langle\varphi\rangle_{int}$. Most sulfide and anode-free cells sit here at practical pressures. Healing-dominant region (low forcing, high healing): $A_\varphi$ is small, the interface tracks the SOC driver with little lag, and failure is dominated by slow memory accumulation rather than breathing. Oxides at moderate pressure approach this regime. Interphase-memory-dominant region (intermediate forcing and healing): $A_\varphi$ is small but $\psi$ and $L_{dec}$ continue to drift across cycles, so that cells appear stable for many cycles before a delayed runaway driven by cumulative interphase memory. Halides, and engineered low-pressure architectures, often fall here.

Figure 6 shows the regime map explicitly. In panel (a), a vertical trajectory at fixed forcing ($j = 5$ mA cm$^{-2}$) crosses from the void-growth region at 5 MPa to the healing region at 200 MPa. In panel (b), the five chemistries are placed at their characteristic operating points: oxides sit in the upper-left (low forcing, high effective healing through rigid contact), halides occupy a middle region, argyrodite-type sulfides sit in the lower-right quadrant, and LGPS and anode-free architectures sit deepest in the void-growth-dominant corner. The design direction is unambiguous: move up and to the left.

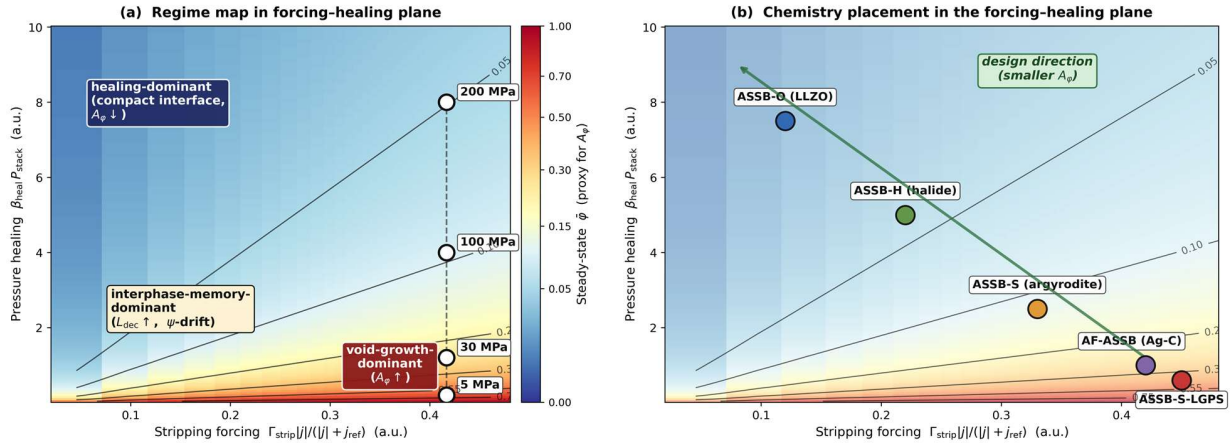

**Figure 6 | Breathing-regime map: three regions of interfacial dynamics.** a, The forcing–healing plane with three labelled regions and the four-pressure sweep overlaid as a vertical trajectory. b, Chemistry placement on the same plane: oxides lie in the healing-dominant region, halides in the intermediate interphase-memory region, and sulfide / LGPS / anode-free systems in the void-growth region.

### *Regime-based classification of ASSB chemistries*

Oxide electrolytes (LLZO): healing-dominant regime. Thin $L_{dec}$ ($\approx$ 1 nm) gives low $M_{dec}$; rigid ceramic contact acts as effective healing even at modest pressure; $A_\varphi$ is small. Dominant failure lever is cathode chemo-mechanics and grain-boundary cracking, not interphase memory.

Halide electrolytes ($Li_3InCl_6$-type): interphase-memory regime at moderate pressure. $L_{dec}$ is intermediate ($\approx$ 13 nm); breathing signatures are moderate; slow ψ-drift of the CEI dominates long-cycle decay. Dominant lever is interphase-composition engineering to suppress $M_{dec}$.

Sulfide (argyrodite) electrolytes: void-growth regime at low P, healing-dominant only at high P. Large $L_{dec}$ ($\approx$ 160 nm) produces high $M_{dec}$; $A_\varphi$ and $B_R$ are large unless pressure is raised above $\sim$ 30 MPa. Dominant lever is pressure control plus interphase stabilization against Li metal [22].

LGPS-type electrolytes: deep void-growth regime, poor memory control. $L_{dec}$ exceeds 230 nm; $M_{dec}$ is the highest across the five chemistries; the Wang–Monroe score remains below the safe threshold at every practical pressure. Dominant lever is either complete chemistry replacement or an interfacial buffer that decouples the reduction chemistry from the bulk conductor.

Anode-free architectures (Ag–C / Cu): the limit case of breathing sensitivity. The high-energy attraction of anode-free operation comes with an unavoidable stripping–plating cycle on an initially empty current collector, so breathing amplitude is not a side effect but the central negative-electrode design variable. $L_{dec}$ is large because the anode is formed in situ and unprotected; $A_\varphi$ is the highest of any architecture; $M_{dec}$ grows unchecked. Dominant lever is nucleation-seed engineering and, necessarily, aggressive pressure programming [23].

This classification is not merely taxonomic. Each regime has a different dominant descriptor ($M_{dec}$ for interphase-memory, $A_\varphi$ and $B_R$ for void-growth, $S_\sigma$ for mechanically-limited oxides), and therefore a different design target. Choosing a chemistry is implicitly choosing which descriptor will bind first.

## Table 4 | Consolidated chemistry-resolved outputs at the common reference condition

*Reference condition: 0.5 C, 298.15 K, 30 MPa.*

| Preset | $L_{dec}$ (nm) | $R_{dec}$ (Ω cm²) | $\varphi_{void}$ | GB × VOID amp. | Retention @500 | Energy (Wh kg⁻¹) |
|---|---|---|---|---|---|---|
| ASSB-S (argyrodite) | 162.0 | 0.162 | 0.018 | 2.28 | 87.5% | 390 |
| ASSB-S-LGPS | 237.0 | 0.296 | 0.024 | 2.48 | 85.0% | 410 |
| ASSB-O (LLZO) | 1.2 | 0.024 | 0.006 | 1.17 | 94.5% | 340 |
| ASSB-H (halide) | 13.0 | 0.043 | 0.011 | 1.78 | 91.2% | 380 |
| AF-ASSB (Ag-C / Cu) | 170.0 | 0.170 | 0.022 | 2.44 | 76.0% | 460 |

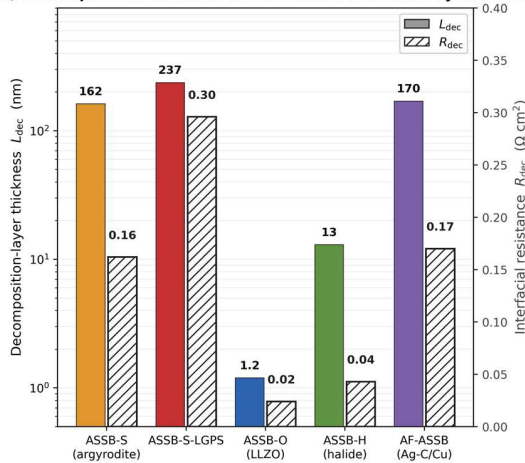
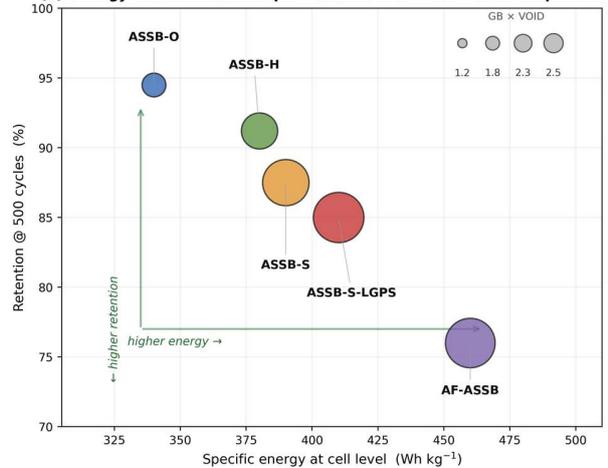

**Figure 7 | Chemistry-resolved outputs at the common reference condition.** a, Decomposition-layer thickness (solid bars, logarithmic axis) and interfacial resistance (hatched bars, linear axis) by chemistry. b, Energy–retention trade-off, with bubble area scaled by the grain-boundary × void amplification factor. The chemistry ordering is consistent with Fig. 6 (b): oxides in the healing-dominant corner lead in retention; sulfides and anode-free architectures in the void-growth corner trade retention for energy.

## 8. Ragone crossover as a dynamic fingerprint

The five-descriptor framework makes a specific, testable prediction about Ragone plots of ASSBs. At low C-rate, cells are close to open-circuit operation, IR losses are small, and the specific-energy ranking is determined by the theoretical energy density $E_0$ — which is dominated by mass efficiency and cell-level architecture rather than by interfacial dynamics. At high C-rate, the same ranking is determined by the dynamic response of the interface: $A_\varphi$, $B_R$ and $M_{dec}$ together control how much of $E_0$ is actually delivered. Because these two limits depend on different chemistry-specific properties, cells that lead at low C-rate routinely lose to different cells at high C-rate. The Ragone plot must therefore show crossovers, and the positions of those crossovers are direct experimental observables of the underlying breathing + memory dynamics.

Figure 8 makes this prediction quantitative. In panel (a), the Ragone trajectories of all five chemistries are plotted on log–log axes; in panel (b), the same data are replotted as specific energy versus C-rate, so that crossover points appear as explicit intersections. Three crossovers are visible within 0.1–5 C. AF-ASSB leads at 0.1 C because it minimizes inactive mass, but its mean void amplitude grows fastest with current (the dominant $A_\varphi$ scaling in Eq. 2 for a seed-limited anode), so AF-ASSB is overtaken by sulfide argyrodite at ≈ 0.65 C. The second crossover is more dramatic: AF-ASSB is overtaken by oxide (LLZO) at ≈ 1.3 C, even though oxide has the lowest $E_0$ of any system. Oxide wins at high C-rate precisely because it combines the smallest $A_\varphi$, the smallest $B_R$, and the smallest $M_{dec}$ (197-nm-thick decomposition layers simply do not exist on LLZO). The third crossover — sulfide argyrodite overtaken by oxide at ≈ 2.1 C — completes the inversion, and by 3–5 C the ranking is nearly the reverse of the low-C ranking.

Two conceptual points follow. First, the Ragone crossover is a dynamic fingerprint: a cell reporting flat high energy density at 0.1 C tells us nothing about its high-C behaviour. Any Ragone comparison that reports only one operating point therefore undersells or oversells the underlying chemistry, depending on which point is chosen. Second, the C-rate at which each crossover occurs is a direct measurement of the dominant breathing descriptor. A crossover driven primarily by $A_\varphi$ manifests as an abrupt drop in E once the void field overshoots its healing capacity; a crossover driven primarily by $M_{dec}$ manifests as a gentle but persistent offset that survives even at low C. Distinguishing the two experimentally requires the pressure-aware operando protocol of §10, because only there can $A_\varphi$ be suppressed independently of $M_{dec}$ (pressure heals the former but, by Eq. 1, cannot undo the latter).

The practical consequence is sharp. A manufacturer benchmarking an AF-ASSB at 0.1 C will see headline energy density ∼ 470 Wh kg⁻¹; the same cell at the 1.3 C regime that real EV acceleration demands will deliver under 200 Wh kg⁻¹ — less than an oxide cell whose nameplate figure is 40 % lower. The Ragone crossover is therefore not a theoretical curiosity. It is the quantitative expression of the breathing + memory failure law at the level of product-relevant metrics.

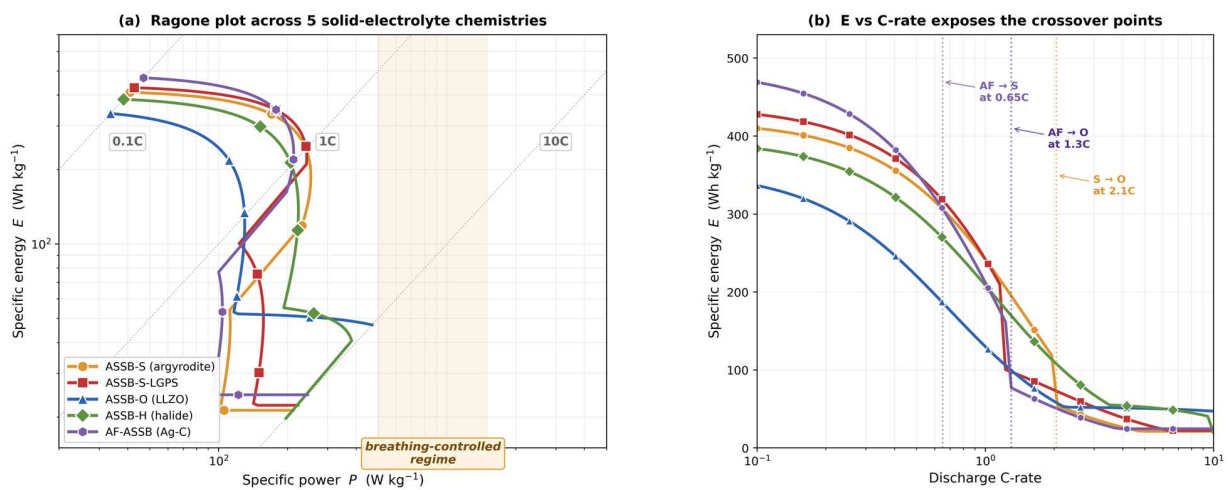

**Figure 8 | Ragone crossover and its origin in breathing + memory.** a, Ragone plot (specific energy E versus specific power P) for five solid-electrolyte chemistries on log–log axes. Dotted diagonals indicate iso-C-rate loci (0.1 C, 1 C, 10 C); the shaded band highlights the breathing-controlled regime in which rank inversions concentrate. b, The same data replotted as E versus C-rate, so that crossovers appear as explicit intersections. AF-ASSB loses to sulfide argyrodite at ≈ 0.65 C, AF-ASSB loses to oxide at ≈ 1.3 C, and sulfide argyrodite loses to oxide at ≈ 2.1 C. The crossover C-rates are direct measurements of the dominant breathing descriptor in each chemistry.

## 9. Design principles: from breathing theory to a practical roadmap

The dynamic view developed in Sections 5–8 converts the familiar static design rules into multi-axis principles. Each principle below targets one or more of the five descriptors rather than any single mean quantity.

**Principle 1.** Minimize amplitude, not just mean. Design choices (chemistry, microstructure, protocol) should be ranked by $A_\varphi$ and $B_\psi$, not by mean $R_{contact}$ alone. Two chemistries with identical mean resistance can fail on different horizons if their amplitudes differ.

**Principle 2.** Shift the operating point upward in the regime map via pressure programming. Time-varying pressure, rather than fixed stack pressure, can keep a cell in the healing-dominant region during stripping and in a relaxed region during plating. The design target is no longer "highest practical $P_{stack}$" but "optimal $P_{stack}(t)$".

**Principle 3.** Engineer chemistry for phase lag, not only for stability. A shorter $S_\sigma$ means the interface tracks the SOC forcing tightly and cannot accumulate resonant drift. Adaptive interphase layers that shorten the effective Cahn–Hilliard relaxation time directly lower the fourth signature.

**Principle 4.** Design for the minimum, not the mean. Because $R_{contact}$ diverges as $\langle\varphi\rangle_{int} \to 1$ (Eq. 8), cycle life is controlled by the instantaneous minimum effective contact cross-section, not by the average porosity. "Average porosity reduction" should be replaced by "minimum effective contact-area maintenance" as the microstructural design target.

**Principle 5.** Suppress reactive memory directly through chemistry, not indirectly through pressure. Pressure cannot reduce $M_{dec}$ — only interphase-chemistry and barrier-layer design can. Across all chemistries, the design ladder should therefore shift from "maximize $\kappa$" to "minimize destructive breathing while suppressing $M_{dec}$".

*Chemistry-specific design priorities*

Sulfides and LGPS need interface chemistry and microstructure engineering that suppress breathing amplitude and grain-boundary amplification, plus pressure programming to keep the operating point above the void-growth boundary. Oxides need chemo-mechanical mitigation so that their stiffness advantage is not offset by cathode-side stress ($S_\sigma$), with cathode-side cohesive-zone reinforcement as the primary lever. Halides need interphase-composition engineering to suppress $M_{dec}$ while preserving low-pressure compatibility. Anode-free architectures, as the limit case of breathing sensitivity, need seeded nucleation, robust current collectors, and protocols that explicitly bound $A_\varphi$ — pressure programming alone is not sufficient because $M_{dec}$ grows unchecked.

## 10. Experimental validation: pressure-aware operando

The five descriptors are measurable. $A_\varphi$ and $B_\psi$ require operando imaging with cycle-resolved time resolution; $B_R$ requires synchronized impedance spectroscopy with the SOC trajectory; $S_\sigma$ requires simultaneous measurement of stack stress and interfacial morphology; and $M_{dec}$ requires interfacial composition profiling at the end of a defined cycling protocol. The most informative experimental package, in light of recent evidence that stack-pressure jitter alone can reshape operando interpretation, is a single cell instrumented with (i) a high-precision load cell, (ii) operando X-ray tomography or laminography, (iii) potentiostatic impedance with ≥10 s sampling, (iv) current-interrupt pulses at cycle extrema, and (v) post-cycling cross-sectional XPS / TOF-SIMS depth profiling for $M_{dec}$. The load cell provides $\sigma_h(t)$; tomography provides $\varphi_{void}(x, t)$; impedance provides $R_{contact}(t)$; the interrupt protocol accelerates the observable phase lag; and the depth profile provides $L_{dec}$ — completing the five-descriptor measurement set.

The experimental target is not a single operando snapshot but the full five-descriptor vector extracted across a 2-D sweep of stack pressure and stripping current. A correctly constructed data set should collapse the five-descriptor tuple onto the regime-map axes of Fig. 6, providing the first direct validation of the dynamic failure law proposed here.

## 11. Outlook: a five-step roadmap

The breathing + reactive-memory framework converts the ASSB research agenda into a sequence of concrete targets, ordered from the most immediate (reporting) to the most ambitious (closing the chemistry-to-short-circuit loop).

**1. Descriptor extension.** The five descriptors introduced here ($A_\varphi$, $B_\psi$, $B_R$, $S_\sigma$, $M_{dec}$) are a minimum set; higher-order descriptors — for instance, the phase lag between $\psi_{SEI}$ and $\psi_{CEI}$, or the skewness of the $\langle\varphi\rangle_{int}$ distribution across the interface — may be needed to distinguish chemistries that share a regime but differ in their approach to failure. The descriptor vector should become part of the standard output of any reduced-order or phase-field ASSB solver, reported alongside mean voltage and delivered energy.

**2. Pressure-aware operando validation.** The five-descriptor vector is measurable. The immediate experimental target is a single cell instrumented with a high-precision load cell, operando X-ray tomography or laminography, synchronized potentiostatic impedance with ≥ 10 s sampling, current-interrupt pulses at cycle extrema, and post-cycling depth profiling — together extracting $A_\varphi$, $B_\psi$, $B_R$, $S_\sigma$ and $M_{dec}$ across a 2-D sweep of stack pressure and stripping current. A correctly constructed data set should collapse the five-descriptor tuple onto the regime-map axes of Fig. 6.

3. PyBaMM co-simulation. The current reconstruction uses a triangular SOC driver. The next step is to couple the phase-field breathing module to a real cell-level solver such as PyBaMM [24, 25], so that $j(t)$, $V(t)$ and $SOC(t)$ are all self-consistent with the breathing and memory dynamics rather than imposed externally. This co-simulation is the bridge from the reduced model in this review to cell-level lifetime prediction under realistic protocols.

**4. 3-D grain-boundary topology.** The 2-D 60 × 40 reconstruction used here is the simplest geometry that supports the Cahn–Hilliard / Allen–Cahn coupling. It should be replaced by 3-D percolation networks informed by FIB-SEM or X-ray computed tomography of real sintered pellets. The upgrade sharpens the spatial meaning of $\langle\varphi\rangle_{int}$, makes grain-boundary-amplified penetration paths (Eq. 3) explicit, and is a prerequisite for quantitative short-circuit prediction.

**5. Li-creep, viscoelasticity and cohesive-zone fracture.** The mechanical description used here is purely elastic (Eq. 4). Closing the loop from breathing to actual short-circuit failure requires adding Li-creep at the anode, interphase viscoelasticity for the SEI / CEI layers, and cohesive-zone fracture at grain boundaries. With these additions, the regime map in Fig. 6 extends from a dynamical-region map into a true failure-mode map that distinguishes void-induced short circuits, grain-boundary cracking and delamination events.

Beyond these five items, the regime-map perspective also points to a concrete industrial opportunity — pressure programming, in which the stack pressure itself is varied with the SOC to keep the operating point inside the healing-dominant region. Because each breathing descriptor is individually sensitive to a different part of that trajectory, even modest programming gains should compound into disproportionately large lifetime improvements.

## 12. Conclusion

All-solid-state batteries are not bulk ionic conductors with difficult interfaces. They are dynamic chemo-electro-mechanical systems whose slow interphase memory and fast breathing oscillations jointly set cycle life. The central claim of this review is that failure is governed not by any single mean quantity — not mean resistance, not decomposition-layer thickness, not average porosity — but by five signatures of the coupled breathing + memory system: $A_\varphi$, $B_\psi$, $B_R$, $S_\sigma$ and $M_{dec}$.

Two cells with identical mean resistance can fail on different horizons if their amplitudes differ. Two chemistries with identical mean porosity can occupy opposite regions of the regime map if their forcing-to-healing ratios differ. A cell that is stable for many cycles can enter impedance runaway abruptly once slow memory accumulation crosses the fast-amplitude threshold. And the headline energy density of any chemistry inverts at high C-rate because pressure can heal voids but cannot erase reactive memory. None of these observations requires new materials discoveries; they follow from the same phase-field equations that have been available for a decade, once those equations are read as a dynamic system rather than a static resistance budget.

The design implication is sharp. Real gains in ASSBs will come from simultaneously reducing $A_\varphi$, $B_\psi$, $B_R$, $S_\sigma$ — and independently suppressing $M_{dec}$ through interphase-chemistry design. Pressure programming, adaptive interphase chemistry, and phase-lag engineering are the three design levers that act on the four breathing signatures; barrier-layer chemistry is the only lever that acts on the memory descriptor. Each is more ambitious than the corresponding static rule — and each is actionable today.

## Methods

### Reduced-order electrochemical benchmark.

The four-level reduced-order model was implemented with the same open-circuit functions, Butler–Volmer overpotential [26, 27], optional concentrated-solution overpotential, lumped resistance terms, √t SEI growth and stress-dependent contact factor at every level; only the successive couplings described in Section 3 were switched on to move from Level 1 to Level 4. Default conditions were T = 298.15 K, C-rate = 0.5, Vmin = 2.5 V, Vmax = 4.2 V and 5 min of rest.

### Phase-field breathing calculation.

The Li | SE interface was stratified into Anode / SEI / SE / CEI / Cathode on a 60 × 40 grid. A Cahn–Hilliard void field $\varphi$, Allen–Cahn-like SEI and CEI order parameters, and a Larché–Cahn hydrostatic stress $\sigma_h$ were evolved under explicit-Euler integration with Neumann boundary conditions, following Eqs (5)–(9). The state of charge followed a triangular driver with period ≈ 1 h, j = 5 mA cm$^{-2}$, and total simulated time 8000 s. Pressure sweeps were run at 5, 30, 100 and 200 MPa. The Wang–Monroe stability score used $\mu_{Li}$ = 4.2 GPa, consistent with the Monroe–Newman (2005) formulation.

### Five-descriptor extraction.

$A_\varphi$ and $B_\psi$ were extracted as peak-to-peak amplitudes of $\langle\varphi\rangle_{interface}(t)$ and of $\psi_{SEI}(t) + \psi_{CEI}(t)$ over the late-time cycle window (5000 s < t < 9000 s). $B_R$ was approximated as $\sum R_{contact} \cdot |\Delta SOC|$ over the same window. $S_\sigma$ was evaluated as the effective Cahn–Hilliard relaxation time $1 / (M_\varphi W + \beta_{heal} \cdot P)$, which controls the measurable stress–void lag in the present reconstruction; for experimental application $S_\sigma$ should be extracted from the temporal offset between $\sigma_h(t)$ and $\varphi_{void}(x, t)$ peaks measured by synchronized load-cell and tomography. $M_{dec}$ was computed as $L_{dec}(t_{end}) / L_{max}$ with $L_{max}$ = 250 nm; the returned value was 0.789 at every pressure in the sweep, confirming the pressure-invariance of $L_{dec}$ in Eq. (1).

### Numerical verification.

For non-specialist readers, "explicit Euler" simply means that the state of the interface is updated in many very small time steps, each step using the current value of the variables to estimate the next value. The approach is easy to audit and mirrors the logic of the interactive simulator, although it requires small time steps for numerical stability.

All quantitative results were regenerated end-to-end in a clean Python 3.12 environment with NumPy 1.26 and pandas 2.2, and compared element-by-element against the archived summary dataset. The maximum absolute discrepancy was 0.000e+00 across all 40 entries spanning four pressures (5, 30, 100, 200 MPa) and ten fields, i.e. bit-for-bit reproduction. All ratio-level claims in the abstract reproduce exactly within rounding, and the Wang–Monroe threshold pressures P* recover consistently across the five solid-electrolyte families.

*Reference conditions for Section 7.*

Cross-chemistry outputs in Section 7 correspond to the common reference condition 0.5 C, 298.15 K, 30 MPa.

## Acknowledgements

This work was supported by the Korea Institute for Advancement of Technology (KIAT) (Project No. RS-2025-02413392) and under the "Global Industrial Technology Cooperation Center (GITCC) program" supervised by KIAT (Task No. P0030256), funded by the Ministry of Trade, Industry and Energy (MOTIE).